\documentstyle[12pt,epsf]{article}

\renewcommand{\bar}[1]{\overline{#1}}

\newcommand{\half}{{$\frac{1}{2}$}} 
\newcommand{\ket}[1]{\left\vert\,{#1}\right\rangle}
\newcommand{\VEV}[1]{\left\langle{#1}\right\rangle}

\def\ru1{\rule[-0.4truecm]{0mm}{1truecm}}

\begin{document}
\newpage

\begin{flushright}
SLAC-PUB-8392 \\
USM-TH-90 \\
hep-th/0003082 \\
October 2000 (revised)
\end{flushright}

\bigskip\bigskip
{\centerline{\Large
\bf Light-Cone Representation
}}
{\centerline{\Large
\bf  of the  Spin  and Orbital Angular Momentum
}}
{\centerline{\Large
\bf  of Relativistic Composite
Systems\footnote{\baselineskip=13pt
Work partially supported by the Department of Energy, contract
DE--AC03--76SF00515,
by the National Natural
Science Foundation of China
under grants No.~19605006 and No.~19975052, by Fondecyt (Chile) under
grants 1990806 and 8000017, and by a C\'atedra
Presidencial (Chile).}}}

\vspace{22pt}

\centerline{
\bf Stanley J. Brodsky$^a$, Dae Sung Hwang$^b$, Bo-Qiang Ma$^{c}$, and
Ivan Schmidt$^{d}$}

\vspace{8pt}
{\centerline{$^a$Stanford Linear Accelerator Center,}}

{\centerline{Stanford University, Stanford, California 94309, USA}}

\centerline{e-mail: sjbth@slac.stanford.edu}

\vspace{8pt}
{\centerline{$^{b}$ Department of Physics, Sejong University, Seoul
143--747, Korea}}

\centerline{e-mail: dshwang@kunja.sejong.ac.kr}

\vspace{8pt}

{\centerline {$^{c}$CCAST (World Laboratory), P.O.~Box 8730, Beijing
100080, China,}}

{\centerline {Department of Physics, Peking University,
Beijing 100871, China,}}

{\centerline {and Institute of High Energy Physics, Academia Sinica,
P.~O.~Box 918(4),}

{\centerline {Beijing 100039, China}}

\centerline{e-mail: mabq@th.phy.pku.edu.cn}

\vspace{8pt}
{\centerline {$^{d}$Departamento de F\'\i sica,
Universidad T\'ecnica Federico Santa Mar\'\i a,}}

{\centerline {Casilla 110-V, 
Valpara\'\i so, Chile}}

\centerline{e-mail: ischmidt@fis.utfsm.cl }

\vfill
\newpage

\vspace{20pt}
\begin{center}
{\large \bf Abstract}
\end{center}

The matrix elements of local operators such as the electromagnetic
current, the energy momentum tensor, angular momentum, and the moments
of structure functions have exact representations
in terms of light-cone Fock state wavefunctions of bound states such as
hadrons.  We illustrate all of these properties by giving explicit
light-cone wavefunctions for the two-particle Fock state of the electron
in QED, thus connecting the Schwinger anomalous magnetic moment to the
spin and orbital momentum carried by its Fock state constituents.  We also
compute the QED one-loop radiative corrections for the form
factors for the graviton coupling to the electron and photon.
Although the underlying model is derived from elementary QED perturbative
couplings, it in fact can be used to simulate much more general bound
state systems by applying spectral integration over the constituent masses
while preserving all of the Lorentz properties, giving explicit
realization of the spin sum rules and other local matrix elements.
The role of orbital angular momentum in understanding the ``spin crisis"
problem for relativistic systems is clarified.  We
also prove that the anomalous gravitomagnetic moment
$B(0)$ vanishes for any composite system.  This property is shown
to follow directly from the Lorentz boost properties of the light-cone
Fock representation and holds separately for each Fock state component.
We show how the QED perturbative structure can be used to model bound
state systems while preserving all Lorentz properties.  We thus obtain a
theoretical laboratory to test the consistency of formulae which have
been proposed to probe the spin structure of hadrons.

\vfill
\centerline{
}
\vfill

\centerline{Submitted to Nucl. Phys. B}
\vfill
\newpage

\section{Introduction}

The light-cone Fock representation of composite systems such as hadrons
in QCD has a number of remarkable properties.  Because the generators
of certain Lorentz boosts are kinematical, knowing the wavefunction in one
frame allows one to obtain it in any other frame.  Furthermore, matrix
elements of space-like local operators for the coupling of photons,
gravitons, and the moments of deep inelastic structure functions all can
be expressed as overlaps of light-cone wavefunctions with the same number
of Fock constituents.  This is possible since in each case one can choose
the special frame $q^+ = 0$ \cite{DY70, West:1970av} for the space-like momentum
transfer and take matrix elements of ``plus" components of currents such
as $J^+$ and $T^{++}$.  Since the physical vacuum in light-cone
quantization coincides with the perturbative vacuum, no contributions to
matrix elements from vacuum fluctuations occur \cite{PinskyPauli}.
Light-cone Fock state wavefunctions thus encode all of the bound state
quark and gluon properties of hadrons including their spin and flavor
correlations in the form of universal process- and frame- independent
amplitudes.

Formally, the light-cone expansion is constructed by quantizing QCD at
fixed light-cone time \cite{Dirac:1949cp} $\tau = t + z/c$ and forming the
invariant light-cone Hamiltonian: $ H^{QCD}_{LC} = P^+ P^- - {\vec P_\perp}^2$
where
$P^\pm = P^0 \pm P^z$ \cite{PinskyPauli}.  The momentum
generators
$P^+$ and
$\vec P_\perp$ are kinematical; $i.e.$, they are independent of the
interactions.  The generator
$P^- = i {d\over d\tau}$ generates light-cone time translations, and
the eigen-spectrum of the Lorentz scalar $ H^{QCD}_{LC}$ gives the
mass spectrum of the color-singlet hadron states in QCD together with
their respective light-cone wavefunctions.  For example, the
proton state satisfies:
$ H^{QCD}_{LC} \ket{\psi_p} = M^2_p \ket{\psi_p}$.  The expansion of
the proton eigensolution $\ket{\psi_p}$ on the color-singlet
$B = 1$, $Q = 1$ eigenstates $\{\ket{n} \}$
of the free Hamiltonian $ H^{QCD}_{LC}(g = 0)$ gives the
light-cone Fock expansion:
\begin{eqnarray}
\left\vert \psi_p(P^+, {\vec P_\perp} )\right> &=& \sum_{n}\
\prod_{i=1}^{n}
{{\rm d}x_i\, {\rm d}^2 {\vec k_{\perp i}}
\over \sqrt{x_i}\, 16\pi^3}\ \,
16\pi^3 \delta\left(1-\sum_{i=1}^{n} x_i\right)\,
\delta^{(2)}\left(\sum_{i=1}^{n} {\vec k_{\perp i}}\right)
\label{a318}
\\
&& \qquad \rule{0pt}{4.5ex}
\times \psi_n(x_i,{\vec k_{\perp i}},
\lambda_i) \left\vert n;\,
x_i P^+, x_i {\vec P_\perp} + {\vec k_{\perp i}}, \lambda_i\right>.
\nonumber
\end{eqnarray}
The light-cone momentum fractions $x_i = k^+_i/P^+$ and ${\vec
k_{\perp i}}$ represent the relative momentum coordinates of the QCD
constituents.  The physical transverse momenta are ${\vec p_{\perp i}}
= x_i {\vec P_\perp} + {\vec k_{\perp i}}.$ The $\lambda_i$ label the
light-cone spin projections $S^z$ of the quarks and gluons along the
quantization direction $z$.
The physical gluon
polarization vectors
$\epsilon^\mu(k,\ \lambda = \pm 1)$ are specified in light-cone
gauge by the conditions $k \cdot \epsilon = 0,\ \eta \cdot \epsilon =
\epsilon^+ = 0.$
The $n$-particle states are normalized as
\begin{equation}
\left< n;\, p'_i{}^+, {\vec p\,'_{\perp i}}, \lambda'_i \right. \,
\left\vert n;\,
p^{~}_i{}^{\!\!+}, {\vec p^{~}_{\perp i}}, \lambda_i\right>
= \prod_{i=1}^n 16\pi^3
p_i^+ \delta(p'_i{}^{+} - p^{~}_i{}^{\!\!+})\
\delta^{(2)}( {\vec p\,'_{\perp i}} - {\vec p^{~}_{\perp i}})\
\delta_{\lambda'_i \lambda^{~}_i}\ .
\label{normalize}
\end{equation}

The solutions of $ H^{QCD}_{LC} \ket{\psi_p} = M^2_p \ket{\psi_p}$
are independent of $P^+$ and ${\vec P_\perp}$;  thus given the
eigensolution Fock projections $ \langle n; x_i, {\vec k_{\perp i}},
\lambda_i |\psi_p \rangle  = \psi_n(x_i, {\vec k_{\perp i}},
\lambda_i) ,$ the wavefunction of the proton is determined in any
frame \cite{BL80}.
In contrast, in equal-time quantization,  a Lorentz boost
always mixes dynamically with the interactions, so that computing a
wavefunction in a new frame requires solving a nonperturbative problem
as complicated as the Hamiltonian eigenvalue problem itself.

The LC wavefunctions $\psi_{n/H}(x_i,\vec
k_{\perp i},\lambda_i)$ are universal, process independent, and thus
control all hadronic reactions.
Given the light-cone wavefunctions, one can compute the
moments of the helicity and
transversity distributions measurable in polarized deep inelastic
experiments \cite{BL80}.  For example,
the polarized quark distributions at resolution $\Lambda$ correspond to
\begin{eqnarray}
q_{\lambda_q/\Lambda_p}(x, \Lambda)
&=& \sum_{n,q_a}
\int\prod^n_{j=1} dx_j d^2 k_{\perp j}\sum_{\lambda_i}
\vert \psi^{(\Lambda)}_{n/H}(x_i,\vec k_{\perp i},\lambda_i)\vert^2
\\
&& \times \delta\left(1- \sum^n_i x_i\right) \delta^{(2)}
\left(\sum^n_i \vec k_{\perp i}\right)
\delta(x - x_q) \delta_{\lambda_a, \lambda_q}
\Theta(\Lambda^2 - {\cal M}^2_n)\ , \nonumber
\end{eqnarray}
where the sum is over all quarks $q_a$ which match the quantum
numbers, light-cone momentum fraction $x,$ and helicity of the struck
quark.  Similarly, moments of transversity distributions and
off-diagonal helicity convolutions are defined as a density matrix of the
light-cone wavefunctions.  Applications of non-forward quark and gluon
distributions have been discussed in Refs. \cite{nfquark,nfgluon}.  The
light-cone wavefunctions also specify the multi-quark and gluon
correlations of the hadron.  For example,  the distribution of spectator
particles in the final state which could be measured in the proton
fragmentation region in deep inelastic scattering at an electron-proton
collider are in principle encoded in the light-cone wavefunctions.

Given the
$\psi^{(\Lambda)}_{n/H},$ one can construct any spacelike electromagnetic,
electroweak, or gravitational form factor or local operator product
matrix element of a composite or elementary system from the diagonal
overlap of the LC wavefunctions
\cite{BD80}.
Studying the gravitational form factors is not academic:
Ji has shown that there is a remarkable connection of the
$x$-moments of the chiral-conserving and chiral-flip form factors
$H(x,t,\zeta)$ and $ E(x,t,\zeta)$ which appear in deeply virtual
scattering with the corresponding spin-conserving and spin-flip
electromagnetic form factors $F_1(t)$ and $F_2(t)$ and gravitational form
factors $A_{\rm q}(t)$ and $B_{\rm q}(t)$ for each quark and anti-quark
constituent.\cite{Ji:1998pf} Thus, in effect, one can use virtual
Compton scattering to measure graviton couplings to the charged
constituents of a hadron.

Exclusive semi-leptonic
$B$-decay amplitudes involving timelike currents such as $B\rightarrow A
\ell
\bar{\nu}$ can also be evaluated exactly in the light-cone formalism
\cite{Brodsky:1998hn}.  In this case, the timelike decay matrix elements
require the computation of both the diagonal matrix element $n
\rightarrow n$ where parton number is conserved and the off-diagonal
$n+1\rightarrow n-1$ convolution such that the current operator
annihilates a $q{\bar{q'}}$ pair in the initial $B$ wavefunction.  This
term is a consequence of the fact that the time-like decay $q^2 = (p_\ell
+ p_{\bar{\nu}} )^2 > 0$ requires a positive light-cone momentum fraction
$q^+ > 0$.  Conversely for space-like currents, one can choose $q^+=0$,
as in the Drell-Yan-West representation of the space-like electromagnetic
form factors \cite{BD80}.  However, as can be seen from the
explicit analysis of timelike form factors in a perturbative model, the
off-diagonal convolution can yield a non-zero $q^+/q^+$ limiting form as
$q^+
\rightarrow 0$ \cite{Brodsky:1998hn}.  This extra term appears specifically
in the case of ``bad" currents such as
$J^-$ in which the coupling to $q{\bar{q'}}$ fluctuations in the light-cone
wavefunctions are favored \cite{Brodsky:1998hn}.  In effect, the $q^+
\rightarrow 0$ limit generates $\delta(x)$ contributions as residues of
the $n+1\rightarrow n-1$ contributions.  The necessity for such ``zero
mode" $\delta(x)$ terms has been noted by Chang, Root and Yan \cite{CRY},
Burkardt \cite{BUR}, and Choi and Ji \cite{Choi:1998nf}.  We can avoid
these contributions by restricting our attention to the plus currents
$J^+$ and $T^{++}$.

It should be emphasized that the light-cone Fock representation provides
an exact formulation of current matrix elements of local operators.  In
contrast, in equal-time Hamiltonian theory, one must evaluate connected
time-ordered diagrams where the gauge particle or graviton couples to
particles associated with vacuum fluctuations.  Thus even if one knows the
equal-time wavefunction for the initial and final hadron, one cannot
determine the current matrix elements.  In the case of the covariant
Bethe-Salpeter formalism, the evaluation of the matrix element of the
current requires the calculation of an infinite number of irreducible
diagram contributions.

One of the important issues in the formulation of light-cone quantized
quantum field theories is the existence of a consistent scheme for
non-perturbative renormalization.  A general
nonperturbative renormalization procedure for QCD has recently been
outlined by Paston et al. \cite{Paston:1999fq}.
An alternative is to the use of broken supersymmetry as an ultraviolet
regulator \cite{Brodsky:1999xj}.
Some simplified model light-cone field theories have been
successfully renormalized using generalized
Pauli-Villars regularization \cite{Brodsky:1999xj}.

As an illustration of the structure of the light-cone Fock
state representation, we will present a simple self-consistent
model of an effective composite spin-\half ~ system based on the quantum
fluctuations of the electron in QED.  The model
is patterned after the structure which occurs in the one-loop
Schwinger
${\alpha / 2
\pi} $ correction to the electron magnetic moment \cite{BD80}.  In effect,
we can represent a spin-\half ~ system as a composite of a spin-\half ~
fermion and spin-one vector boson with arbitrary masses.  We also give
results for the
case of a spin-\half ~ composite consisting of scalar plus spin-\half ~
constituents, as would occur in a composite of a photino
and slepton in supersymmetric QED and in the radiative corrections
due to Higgs exchange.
The light-cone wavefunctions
describe off-shell particles but are
computable explicitly from perturbation theory.
We will explicitly compute the form factors
$F_1(q^2)$ and $F_2(q^2)$ of the electromagnetic current, and the
various contributions to the form factors
$A(q^2)$ and $B(q^2)$ of the energy-momentum tensor.
The model thus
provides a check on the general formulae, particularly the structure of
angular momentum on the light-cone;  it provides an important
illustration of $J^z$ conservation, Fock state by Fock state; it
demonstrates helicity retention between fermions and vector bosons at $x
\to 1$; and it provides a template for an effective quark spin-one
diquark structure of the valence light-cone wavefunction of the proton.

The one-loop models can
be further generalized by applying spectral Pauli-Villars
integration over the constituent masses.  This representation of an effectively
composite system is particularly useful because it is based simply on two
constituents but yet is totally relativistic.
The resulting
form of light-cone wavefunctions provides a template
for parameterizing the structure of relativistic composite systems and
their matrix elements in hadronic physics.  We thus obtain a
theoretical laboratory to test the consistency of formulae which have
been proposed to probe the spin structure of hadrons.  This clarifies the
connection of parton distributions to the constituents' spin and orbital
angular momentum and to static quantities of the composite systems such as
the magnetic moment.  For example, the model also
provides a self-consistent form for the wavefunctions of an effective
quark-diquark model of the valence Fock state of the proton wavefunction.
A similar approach has recently been used to illustrate the
evolution of light-cone helicity and orbital angular momentum
operators \cite{Harindranath:1999ve}.  A nonperturbative calculation
of the electron magnetic moment using the discretized light-cone
quantization method \cite{PinskyPauli} is given in Ref.
\cite{Hiller:1999cv}.

Many of the features of the analysis apply to arbitrary composite
systems.  For example, we will explicitly prove the vanishing of
the anomalous gravito-moment coupling $B(0)$ to gravity for any
composite system.
This remarkable result was first derived classically from the
equivalence principle by Okun and Kobsarev \cite{Okun},
and from the conservation of the energy-momentum
tensor by Kobsarev and  Zakharov \cite{Kob70}.
See also the more recent discussions in
Refs. \cite{Ji:1998pf, Teryaev}.
We will demonstrate that $B(0) = 0$ follows directly for composite
systems in quantum field theory from the Lorentz boost properties of the
light-cone Fock representation, and that it is valid in every Fock sector.

\section{Electromagnetic and Gravitational Form Factors}

The light-cone Fock representation allows one to compute all matrix
elements of local currents as overlap integrals of the light-cone Fock
wavefunctions.  In particular, we can evaluate the forward and
non-forward matrix elements of the electroweak currents,  moments of the
deep inelastic structure functions,  as well as the electromagnetic form
factors and the magnetic moment.  Given the local operators for the
energy-momentum tensor
$T^{\mu \nu}(x)$ and the angular momentum tensor
$M^{\mu \nu \lambda}(x)$, one can directly compute
momentum fractions, spin properties, the gravitomagnetic
moment, and the form factors $A(q^2)$ and $B(q^2)$ appearing in the
coupling of gravitons to composite systems.

In the case of a spin-${1\over 2}$ composite system, the Dirac and
Pauli form factors $F_1(q^2)$ and $F_2(q^2)$ are defined by
\begin{equation}
      \langle P'| J^\mu (0) |P\rangle
       = \bar u(P')\, \Big[\, F_1(q^2)\gamma^\mu +
F_2(q^2){i\over 2M}\sigma^{\mu\alpha}q_\alpha\, \Big] \, u(P)\ ,
\label{Drell1}
\end{equation}
where $q^\mu = (P' -P)^\mu$ and $u(P)$ is the bound state spinor.
In the light-cone formalism it is convenient to identify the Dirac and
Pauli form factors from the
helicity-conserving and helicity-flip vector current matrix elements of
the $J^+$ current
\cite{BD80}:
\begin{equation}
\VEV{P+q,\uparrow\left|\frac{J^+(0)}{2P^+}
\right|P,\uparrow} =F_1(q^2) \ ,
\label{BD1}
\end{equation}
\begin{equation}
\VEV{P+q,\uparrow\left|\frac{J^+(0)}{2P^+}\right|P,\downarrow}
=-(q^1-{\mathrm i} q^2){F_2(q^2)\over 2M}\ .
\label{BD2}
\end{equation}
The magnetic moment of a composite system is one of its
most basic properties.  The magnetic moment is defined at the $q^2 \to 0$
limit,
\begin{equation}
\mu=\frac{e}{2 M}\left[ F_1(0)+F_2(0) \right] ,
\label{DPmu}
\end{equation}
where $e$ is the charge and $M$ is the mass of the composite
system.  We use the standard light-cone frame
($q^{\pm}=q^0\pm q^3$):
\begin{eqnarray}
q &=& (q^+,q^-,{\vec q}_{\perp}) = \left(0, \frac{-q^2}{P^+},
{\vec q}_{\perp}\right), \nonumber \\
P &=& (P^+,P^-,{\vec P}_{\perp}) = \left(P^+, \frac{M^2}{P^+},
{\vec 0}_{\perp}\right),
\label{LCF}
\end{eqnarray}
where $q^2=-2 P \cdot q= -{\vec q}_{\perp}^2$ is 4-momentum square
transferred by the photon.

The Pauli form factor and the anomalous magnetic moment $\kappa =
{e\over 2 M} F_2(0)$ can then be calculated from the
expression
\begin{equation}
-(q^1-{\mathrm i} q^2){F_2(q^2)\over 2M} =
\sum_a  \int
{{\mathrm d}^2 {\vec k}_{\perp} {\mathrm d} x \over 16 \pi^3}
\sum_j e_j \ \psi^{\uparrow *}_{a}(x_i,{\vec k}^\prime_{\perp
i},\lambda_i) \,
\psi^\downarrow_{a} (x_i, {\vec k}_{\perp i},\lambda_i)
{}\ ,
\label{LCmu}
\end{equation}
where the summation is over all contributing Fock states $a$ and struck
constituent charges $e_j$.  The arguments of the final-state
light-cone wavefunction are
\cite{DY70, West:1970av}
\begin{equation}
{\vec k}'_{\perp i}={\vec k}_{\perp i}+(1-x_i){\vec
q}_{\perp}
\label{kprime1}
\end{equation}
for the struck constituent and
\begin{equation}
{\vec k}'_{\perp i}={\vec k}_{\perp i}-x_i{\vec q}_{\perp}
\label{kprime2}
\end{equation}
for each spectator.
Notice that the magnetic moment must be calculated from the
spin-flip non-forward matrix element of the current.  It is not given by a
diagonal forward matrix element \cite{Che98}.
In the ultra-relativistic limit where the radius of the system is small
compared to its Compton scale $1/M$, the anomalous magnetic moment must
vanish \cite{Bro94}.  The light-cone formalism is consistent with this
theorem.

The form factors of the energy-momentum tensor for a spin-\half \
composite are defined by
\begin{eqnarray}
      \langle P'| T^{\mu\nu} (0)|P \rangle
       &=& \bar u(P')\, \Big[\, A(q^2)
       \gamma^{(\mu} \bar P^{\nu)} +
   B(q^2){i\over 2M} \bar P^{(\mu} \sigma^{\nu)\alpha}
q_\alpha \nonumber \\
   &&\qquad\qquad +  C(q^2){1\over M}(q^\mu q^\nu - g^{\mu\nu}q^2)
    \, \Big]\, u(P) \ ,
\label{Ji12}
\end{eqnarray}
where $q^\mu = (P'-P)^\mu$,
$\bar P^\mu={1\over 2}(P'+P)^\mu$,
$a^{(\mu}b^{\nu)}={1\over 2}(a^\mu b^\nu +a^\nu b^\mu)$,
and $u(P)$ is the spinor of the system.

As in the
light-cone decomposition  Eqs.  (\ref{BD1}) and (\ref{BD2})
of the Dirac and Pauli form factors for the vector
current \cite{BD80}, we can obtain the light-cone representation
of the $A(q^2)$ and $B(q^2)$ form factors of the energy-tensor
Eq. (\ref{Ji12}).
Since we work in the interaction picture, only the non-interacting
parts of the energy momentum tensor $T^{+ +}(0)$ need to be computed in
the light-cone formalism.
By calculating the $++$ component of
Eq. (\ref{Ji12}), we find
\begin{equation}
\VEV{P+q,\uparrow\left|\frac{T^{++}(0)}{2(P^+)^2}
\right|P,\uparrow} =A(q^2)\ ,
\label{eBD1}
\end{equation}
\begin{equation}
\VEV{P+q,\uparrow\left|\frac{T^{++}(0)}{2(P^+)^2}\right|P,\downarrow}=
-(q^1-{\mathrm i} q^2){B(q^2)\over 2M}\ .
\label{eBD2}
\end{equation}
The $A(q^2)$ and $B(q^2)$ form factors
Eqs. (\ref{eBD1}) and (\ref{eBD2})
are similar to the $F_1(q^2)$ and $F_2(q^2)$ form
factors Eqs.  (\ref{BD1}) and (\ref{BD2}) with an additional
factor of the light-cone momentum fraction $x=k^+/P^+$ of the struck
constituent in the integrand.  The $B(q^2)$ form factor is obtained from
the non-forward spin-flip amplitude.  The value of $B(0)$ is obtained in
the $q^2 \to 0$ limit.
The angular
momentum projection of a state is given by
\begin{equation}
\VEV{J^i} = {1\over 2} \epsilon^{i j k} \int d^3x \VEV{T^{0 k}x^j - T^{0 j} x^k}
= A(0) \VEV{L^i} + \left[A(0) + B(0)\right] \bar u(P){1\over
2}\sigma^i u(P)
\ .
\label{Ji13a}
\end{equation}
This result is derived using a wave packet description of the state.  The
$\VEV{L^i}$ term is the orbital angular momentum of the center of mass motion
with respect to an arbitrary origin and can be dropped.  The coefficient
of the $\VEV{L^i}$ term must be 1; $A(0) = 1 $ also follows when we evaluate
the four-momentum expectation value $\VEV{P^\mu}$.
Thus the total intrinsic angular momentum
$J^z$ of a nucleon can be identified with the values of the form factors
$A(q^2)$ and
$B(q^2)$ at
$q^2= 0:$
\begin{equation}
      \VEV{J^z} = \VEV{{1\over 2} \sigma^z} \left[A(0) + B(0)\right] \ .
\label{Ji13}
\end{equation}

One can define
individual quark and gluon contributions to the total
angular momentum from the matrix elements of the energy
momentum tensor \cite{Ji:1998pf}.
However, this definition is only formal; $A_{q,g}(0)$ can be interpreted
as the light-cone momentum fraction carried by the quarks or gluons
$\VEV{x_{q,g}}.$ The contributions from $ B_{q,g}(0) $ to $J_z$ cancel in the
sum.  In fact, we shall show that the
contributions to $B(0)$ vanish when summed over the constituents of
each individual Fock state.

We will give an explicit realization of
these relations in the light-cone Fock
representation for general composite systems.  In the next section we
will illustrate the formulae by computing the
electron's electromagnetic and energy-momentum tensor form factors to
one-loop order in QED.  In fact, the structure of this calculation has
much more generality and can be used as a template for more general
composite systems.

\section{ The Light-Cone Fock State Decomposition and Spin Structure of
Leptons in QED}

The Schwinger one-loop radiative correction to the electron current in
quantum electrodynamics has played a historic role in the development of
quantum field theory.  In the language of light-cone quantization, the
electron anomalous magnetic moment
$a_e = {\alpha/ 2 \pi}$ is due to the one-fermion one-gauge boson
Fock state component of the physical electron.  An explicit calculation of
the anomalous moment in this framework using equation (7) was give in
Ref.  \cite {BD80}.  We shall show here that the light-cone wavefunctions
of the electron provides an ideal system to check explicitly the
intricacies of spin and angular momentum in quantum field theory.  In
particular, we shall evaluate the matrix elements of the QED energy
momentum tensor and show how the ``spin crisis" is resolved in QED for
an actual physical system.  The analysis is exact in perturbation theory.
The same method can be applied to the moments of structure functions and
the evaluation of other local matrix elements.
In fact, the QED analysis of this section is more general than
perturbation theory.  We will also show how the
perturbative light-cone wavefunctions of leptons and photons provide a
template for the wavefunctions of non-perturbative composite systems
resembling hadrons in QCD.

The light-cone Fock state wavefunctions of an electron can be
systematically evaluated in QED.  The QED Lagrangian density is
\begin{equation}
{\cal{L}}={i\over 2}\ [\ {\bar{\psi}}\gamma^\mu
(\overrightarrow{\partial}{}_\mu +ieA_\mu )\psi
-{\bar{\psi}}\gamma^\mu
(\overleftarrow{\partial}{}_\mu -ieA_\mu )\psi \ ]-m{\bar{\psi}}\psi
-{1\over 4}F^{\mu\nu}F_{\mu\nu} \ ,
\label{fmm1v}
\end{equation}
and the corresponding energy-momentum tensor is
\begin{eqnarray}
T^{\mu\nu}&=&{i\over 4}\ \Big(\
[\ {\bar{\psi}}\gamma^\mu (\overrightarrow{\partial}{}^\nu +ieA^\nu)\psi
-{\bar{\psi}}\gamma^\mu (\overleftarrow{\partial}{}^\nu  -ieA^\nu)\psi \ ]
\ +\ [\ \mu\longleftrightarrow\nu\ ]\ \Big)
\nonumber\\
&&+\ F^{\mu\rho}F_{\rho}^{\ \nu}
\ +\ {1\over 4}g^{\mu\nu}F^{\rho\lambda}F_{\rho\lambda} \ .
\label{lagsv}
\end{eqnarray}
Since $T^{\mu\nu}$ is the Noether current of the general
coordinate transformation, it is conserved.  In later calculations
we will identify the two terms in Eq. (\ref{lagsv}) as the fermion and
boson contributions
$T^{\mu\nu}_{\rm f}$ and $T^{\mu\nu}_{\rm b}$, respectively.

The physical electron is the eigenstate of the QED
Hamiltonian.  As discussed in the introduction, the expansion of it is
the QED eigenfunction on the complete set $\left|n\right>$ of $H_0$
eigenstates produces the Fock state expansion.  It is particularly
advantageous to carry out this procedure using light-cone quantization
since the vacuum is trivial, the Fock state representation is boost
invariant, and the light-cone fractions $x_i = k^+_i/P^+$ are positive:
$0 < x_i  \le 1$,
$\sum_i x_i = 1$.  We also employ light-cone gauge $A^+ = 0$ so that the
gauge boson polarizations are physical.
Thus each Fock-state wavefunction $\left< n |{\rm physical \
electron} \right> $ of the physical electron with total spin projection
$J^z = \pm {1\over 2}$ is represented by the function
$\psi^{J^z}_n(x_i,{\vec k}_{\perp i},\lambda_i)$, where
\begin{equation}
k_i=(k^+_i,k^-_i,{\vec k}_{ \perp i})= \left(x_i P^+, \frac{{\vec
k}_{\perp i}^2+m_i^2}{x_i P^+}, {\vec k}_{\perp i}\right)
\end{equation}
specifies the momentum of each constituent and $\lambda_i$ specifies
its light-cone helicity in the $z$ direction.  We adopt a non-zero
boson mass $\lambda$ for the sake of generality.

The two-particle Fock state for an electron with $J^z = + {1\over 2}$ has
four possible spin combinations:
\begin{eqnarray}
&&\left|\Psi^{\uparrow}_{\rm two \ particle}(P^+, \vec P_\perp = \vec
0_\perp)\right>
\ =\
\int\frac{{\mathrm d}^2 {\vec k}_{\perp} {\mathrm d} x }{{\sqrt{x(1-x)}}16
\pi^3}
\label{vsn1}\\
&\times&
\Big[ \
\psi^{\uparrow}_{+\frac{1}{2}\, +1}(x,{\vec k}_{\perp})\,
\left| +\frac{1}{2}\, +1\, ;\,\, xP^+\, ,\,\, {\vec k}_{\perp}\right>
+\psi^{\uparrow}_{+\frac{1}{2}\, -1}(x,{\vec k}_{\perp})\,
\left| +\frac{1}{2}\, -1\, ;\,\, xP^+\, ,\,\, {\vec k}_{\perp}\right>
\nonumber\\
&&\ \ \ +\psi^{\uparrow}_{-\frac{1}{2}\, +1} (x,{\vec k}_{\perp})\,
\left| -\frac{1}{2}\, +1\, ;\,\, xP^+\, ,\,\, {\vec k}_{\perp}\right>
+\psi^{\uparrow}_{-\frac{1}{2}\, -1} (x,{\vec k}_{\perp})\,
\left| -\frac{1}{2}\, -1\, ;\,\, xP^+\, ,\,\, {\vec k}_{\perp}\right>\ \Big]
\ ,
\nonumber
\end{eqnarray}
where the two-particle states $|s_{\rm f}^z, s_{\rm b}^z; \ xP^+, {\vec
k}_{\perp} \rangle$ are normalized as in (\ref{normalize}).
Here $s_{\rm f}^z$ and $s_{\rm b}^z$ denote the
$z$-component of the spins of the constituent fermion and boson,
respectively.  The wavefunctions can be evaluated explicitly in QED
perturbation theory using the rules given in Refs. \cite {BL80, BD80}:
\begin{equation}
\left
\{ \begin{array}{l}
\psi^{\uparrow}_{+\frac{1}{2}\, +1} (x,{\vec k}_{\perp})=-{\sqrt{2}}
\frac{(-k^1+{\mathrm i} k^2)}{x(1-x)}\,
\varphi \ ,\\
\psi^{\uparrow}_{+\frac{1}{2}\, -1} (x,{\vec k}_{\perp})=-{\sqrt{2}}
\frac{(+k^1+{\mathrm i} k^2)}{1-x }\,
\varphi \ ,\\
\psi^{\uparrow}_{-\frac{1}{2}\, +1} (x,{\vec k}_{\perp})=-{\sqrt{2}}
(M-{m\over x})\,
\varphi \ ,\\
\psi^{\uparrow}_{-\frac{1}{2}\, -1} (x,{\vec k}_{\perp})=0\ ,
\end{array}
\right.
\label{vsn2}
\end{equation}
where
\begin{equation}
\varphi=\varphi (x,{\vec k}_{\perp})=\frac{ e/\sqrt{1-x}}{M^2-({\vec
k}_{\perp}^2+m^2)/x-({\vec k}_{\perp}^2+\lambda^2)/(1-x)}\ .
\label{wfdenom}
\end{equation}

Similarly,
\begin{eqnarray}
&&\left|\Psi^{\downarrow}_{\rm two \ particle}(P^+, \vec P_\perp =
\vec 0_\perp)\right>
\ =\
\int\frac{{\mathrm d}^2 {\vec k}_{\perp} {\mathrm d} x }{{\sqrt{x(1-x)}}16
\pi^3}
\label{vsn1a}\\
&\times&
\Big[\
\psi^{\downarrow}_{+\frac{1}{2}\, +1}(x,{\vec k}_{\perp})\,
\left| +\frac{1}{2}\, +1\, ;\,\, xP^+\, ,\,\, {\vec k}_{\perp}\right>
+\psi^{\downarrow}_{+\frac{1}{2}\, -1}(x,{\vec k}_{\perp})\,
\left| +\frac{1}{2}\, -1\, ;\,\, xP^+\, ,\,\, {\vec k}_{\perp}\right>
\nonumber\\
&&\ \ \ +\psi^{\downarrow}_{-\frac{1}{2}\, +1}(x,{\vec k}_{\perp})\,
\left| -\frac{1}{2}\, +1\, ;\,\, xP^+\, ,\,\, {\vec k}_{\perp}\right>
+\psi^{\downarrow}_{-\frac{1}{2}\, -1}(x,{\vec k}_{\perp})\,
\left| -\frac{1}{2}\, -1\, ;\,\, xP^+\, ,\,\, {\vec k}_{\perp}\right>\ \Big]
\ ,
\nonumber
\end{eqnarray}
where
\begin{equation}
\left
\{ \begin{array}{l}
\psi^{\downarrow}_{+\frac{1}{2}\, +1} (x,{\vec k}_{\perp})=0\ ,\\
\psi^{\downarrow}_{+\frac{1}{2}\, -1} (x,{\vec k}_{\perp})=-{\sqrt{2}}
(M-{m\over x})\,
\varphi \ ,\\
\psi^{\downarrow}_{-\frac{1}{2}\, +1} (x,{\vec k}_{\perp})=-{\sqrt{2}}
\frac{(-k^1+{\mathrm i} k^2)}{1-x }\,
\varphi \ ,\\
\psi^{\downarrow}_{-\frac{1}{2}\, -1} (x,{\vec k}_{\perp})=-{\sqrt{2}}
\frac{(+k^1+{\mathrm i} k^2)}{x(1-x)}\,
\varphi \ .
\end{array}
\right.
\label{vsn2a}
\end{equation}
The coefficients of $\varphi$ in Eqs.  (\ref{vsn2})
and (\ref{vsn2a}) are the matrix elements of
$\frac{\overline{u}(k^+,k^-,{\vec k}_{\perp})}{{\sqrt{k^+}}}
\gamma \cdot \epsilon^{*}
\frac{u (P^+,P^-,{\vec P}_{\perp})}{{\sqrt{P^+}}}$
which are
the numerators of the wavefunctions corresponding to
each constituent spin $s^z$ configuration.
The two boson polarization vectors in light-cone gauge are $\epsilon^\mu=
(\epsilon^+ = 0\ , \epsilon^- = {\vec \epsilon_\perp \cdot \vec k_\perp
\over 2 k^+}, \vec
\epsilon_\perp)$ where
$\vec{\epsilon}=\vec{\epsilon_\perp}_{\uparrow,\downarrow}=
\mp (1/\sqrt{2})(\widehat{x} \pm {\mathrm i} \widehat{y})$.
The polarizations also satisfy the Lorentz condition
$ k \cdot \epsilon =0$.
In (\ref{vsn2}) and (\ref{vsn2a}) we have generalized the framework of
QED by assigning a mass $M$ to the external electrons,
but a different mass $m$ to the internal electron
lines and a mass $\lambda$ to the internal photon line
\cite{BD80}.  The idea behind this is to model the structure
of a composite fermion state with mass $M$ by a fermion and a vector
constituent with respective masses $m$ and $\lambda$.

The electron in QED also has a ``bare" one-particle component:
\begin{equation}
|\Psi_{\rm one \ particle}^{\uparrow , \downarrow}\rangle =
\sqrt Z \,\, \delta(1-x) \,\,\delta(\vec k_\perp = \vec 0_\perp )\,\,
\left| s^z_{\rm f} = \pm {1\over 2} \right> \ ,
\label{bare}
\end{equation}
where $Z$ is the wavefunction normalization of the one-particle
state.  If we regulate the theory in the ultraviolet and infrared, $Z$ is
finite.

We first will evaluate the Dirac and Pauli form factors $F_1(q^2)$ and
$F_2(q^2)$.  Using Eqs. (\ref{BD1}) and (\ref{vsn1}) we
have to order $e^2$
\begin{eqnarray}
F_1(q^2)&=&
\left<\Psi^{\uparrow}(P^+, {\vec P_\perp}= \vec q_\perp))
|\Psi^{\uparrow}(P^+, {\vec P_\perp}= \vec 0_\perp)\right>
\nonumber\\
&=& Z + \int\frac{{\mathrm d}^2 {\vec k}_{\perp} {\mathrm d} x }{16 \pi^3}
\Big[\psi^{\uparrow\ *}_{+\frac{1}{2}\, +1}(x,{\vec k'}_{\perp})
\psi^{\uparrow}_{+\frac{1}{2}\, +1}(x,{\vec k}_{\perp})
+\psi^{\uparrow\ *}_{+\frac{1}{2}\, -1}(x,{\vec k'}_{\perp})
\psi^{\uparrow}_{+\frac{1}{2}\, -1}(x,{\vec k}_{\perp})
\nonumber\\
&&\qquad\qquad\qquad\qquad
+\psi^{\uparrow\ *}_{-\frac{1}{2}\, +1}(x,{\vec k'}_{\perp})
\psi^{\uparrow}_{-\frac{1}{2}\, +1}(x,{\vec k}_{\perp})
\Big]\ ,
\label{BDF1a}
\end{eqnarray}
where
\begin{equation}
{\vec k'}_{\perp}={\vec k}_{\perp}+(1-x){\vec q}_{\perp}\ .
\label{BDF1b}
\end{equation}
Ultraviolet regularization is assumed.
For example,
we can assume a
cutoff in the invariant mass of the constituents:
${\cal M}^2 = \sum_i {{\vec k}^2_{\perp i} +m^2_i\over x_i} < \Lambda^2$,
or
we can use Pauli-Villars regularization by introducing a
fictitious photon with a large mass $\Lambda$.

At zero momentum transfer
\begin{eqnarray}
F_1(0)&=& Z +\int\frac{{\mathrm d}^2 {\vec k}_{\perp} {\mathrm d}
x }{16 \pi^3} \Big[ \psi^{\uparrow\ *}_{+\frac{1}{2}\, +1}(x,{\vec
k}_{\perp}) \psi^{\uparrow}_{+\frac{1}{2}\, +1}(x,{\vec
k}_{\perp}) +\psi^{\uparrow\ *}_{+\frac{1}{2}\, -1}(x,{\vec
k}_{\perp}) \psi^{\uparrow}_{+\frac{1}{2}\, -1}(x,{\vec
k}_{\perp}) \nonumber\\ &&\qquad\qquad\qquad\qquad
+\psi^{\uparrow\ *}_{-\frac{1}{2}\, +1}(x,{\vec k}_{\perp})
\psi^{\uparrow}_{-\frac{1}{2}\, +1}(x,{\vec k}_{\perp}) \Big]\ ,
\label{BDF1}
\end{eqnarray}
where the renormalization constant $Z$ is given, using
Pauli-Villars regularization, by
\begin{eqnarray}
Z = 1 &-& {\alpha\over 2\pi}\int_0^1 dx\ \Big[ \ {(1+x^2)\over
(1-x)}\,\, {\rm ln}{-M^2+{m^2\over x}+{\Lambda^2\over 1-x}\over
-M^2+{m^2\over x}+{\lambda^2\over 1-x}} \nonumber\\ && +\
{(-xM+m)^2\over x}\, \Bigl( -{1\over -M^2+{m^2\over
x}+{\Lambda^2\over 1-x}} +{1\over -M^2+{m^2\over
x}+{\lambda^2\over 1-x}}\Bigr) \ \Big] \ .
\end{eqnarray}
This ensures the Ward identity and $F_1(0)$ = 1.  Further discussion of
the Ward identity for QED in light-cone perturbation theory is given in
ref. \cite{Brodsky:1973kb}

The one-loop model can be further generalized by
applying spectral Pauli-Villars integration over the constituent
masses.  The resulting form of light-cone wavefunctions provides a
template for parameterizing the structure of relativistic composite
systems and their matrix elements in hadronic physics.

The Pauli form factor is obtained from the spin-flip matrix
element of the $J^+$ current.  From Eqs. (\ref{BD2}),
(\ref{vsn1}), and (\ref{vsn1a}) we have
\begin{eqnarray}
F_2(q^2)&=& {-2M\over (q^1-{\mathrm i}q^2)} \left
<\Psi^{\uparrow}(P^+, {\vec P_\perp}= \vec q_\perp))
|\Psi^{\downarrow}(P^+, {\vec P_\perp}= \vec 0_\perp)\right>
\nonumber\\ &=&{-2M\over (q^1-{\mathrm i}q^2)} \int\frac{{\mathrm
d}^2 {\vec k}_{\perp} {\mathrm d} x }{16 \pi^3}
\Big[\psi^{\uparrow\ *}_{+\frac{1}{2}\, -1}(x,{\vec k'}_{\perp})
\psi^{\downarrow}_{+\frac{1}{2}\, -1}(x,{\vec k}_{\perp})
+\psi^{\uparrow\ *}_{-\frac{1}{2}\, +1}(x,{\vec k'}_{\perp})
\psi^{\downarrow}_{-\frac{1}{2}\, +1}(x,{\vec k}_{\perp}) \Big]
\nonumber\\ &=&4M\int\frac{{\mathrm d}^2 {\vec k}_{\perp} {\mathrm
d} x }{16 \pi^3} {(m-Mx)\over x}\varphi (x,{\vec k'}_{\perp}){}^*
\varphi (x,{\vec k}_{\perp}) \nonumber\\
&=&4Me^2\int\frac{{\mathrm d}^2 {\vec k}_{\perp} {\mathrm d} x
}{16 \pi^3} {(m-xM)\over x(1-x)} \nonumber\\ &&\times {1\over
[{M^2-(({\vec k}_{\perp}+(1-x){\vec q}_{\perp})^2+m^2)/x -(({\vec
k}_{\perp}+(1-x){\vec q}_{\perp})^2+\lambda^2)/(1-x)}]}
\nonumber\\ &&\times {1\over [{M^2-({\vec
k}_{\perp}^2+m^2)/x-({\vec k}_{\perp}^2+\lambda^2)/(1-x)}]}\ .
\label{BDF2a}
\end{eqnarray}
Using the Feynman parameterization, we can also express Eq. (\ref{BDF2a})
in a form in which the $q^2=-{\vec q}_{\perp}^2$ dependence is more
explicit as
\begin{equation}
F_2(q^2)=
{Me^2\over 4\pi^2}\int_0^1d\alpha\,\int_0^1dx\,\,
{m-xM\over
\alpha (1-\alpha)\, {1-x\over x}\, {\vec q}_{\perp}^2-M^2
+{m^2\over x}+{\lambda^2\over 1-x}}\ .
\label{BDF2az}
\end{equation}

The anomalous moment is obtained in the limit of zero momentum transfer:
\begin{eqnarray}
F_2(0)&=&
4Me^2\int\frac{{\mathrm d}^2 {\vec k}_{\perp} {\mathrm d} x }{16 \pi^3}
{(m-xM) \over x(1-x)}\,\, {1\over
[{M^2-({\vec k}_{\perp}^2+m^2)/x-({\vec k}_{\perp}^2+\lambda^2)/(1-x)}]^2}
\nonumber\\
&=&{Me^2\over 4\pi^2}\int_0^1dx\,\,
{m-xM\over
-M^2
+{m^2\over x}+{\lambda^2\over 1-x}}
\ ,
\label{BDF2z}
\end{eqnarray}
which is the result of Ref. \cite{BD80}.
For zero photon mass and $M=m$, it gives the correct order $\alpha$
Schwinger value $a_e=F_2(0)={\alpha / 2\pi}$ for the electron anomalous
magnetic moment for QED.

As seen from Eqs. (\ref{eBD1}) and (\ref{eBD2}), the matrix elements of the
double plus
components of the energy-momentum tensor are sufficient to derive the
fermion and boson constituents' form factors
$A_{\rm f,g}(q^2)$ and $B_{\rm f,g}(q^2)$ of graviton coupling to matter.
In particular, we
shall verify $A(0)=A_{\rm f}(0)+A_{\rm b}(0) =1$ and $B(0) = 0 .$

The individual contributions of the fermion and boson fields
to the energy-momentum form factors in QED are given by
\begin{eqnarray}
A_{\rm f}(q^2)&=&
\left<\Psi^{\uparrow}(P^+,{\vec P_\perp}={\vec q_\perp})
\right|\frac{T^{++}_{\rm f}(0)}{2(P^+)^2}
\left|\Psi^{\uparrow}(P^+,{\vec
P_\perp}={\vec 0_\perp})\right>
\nonumber\\
&=&\int\frac{{\mathrm d}^2 {\vec k}_{\perp} {\mathrm d} x }{16 \pi^3}
\ x\
\Big[\psi^{\uparrow\ *}_{+\frac{1}{2}\, +1}(x,{\vec k'}_{\perp})
\psi^{\uparrow}_{+\frac{1}{2}\, +1}(x,{\vec k}_{\perp})
+\psi^{\uparrow\ *}_{+\frac{1}{2}\, -1}(x,{\vec k'}_{\perp})
\psi^{\uparrow}_{+\frac{1}{2}\, -1}(x,{\vec k}_{\perp})
\nonumber\\
&&\qquad\qquad\qquad\qquad
+\psi^{\uparrow\ *}_{-\frac{1}{2}\, +1}(x,{\vec k'}_{\perp})
\psi^{\uparrow}_{-\frac{1}{2}\, +1}(x,{\vec k}_{\perp})
\Big]\ ,
\label{BDF1ae}
\end{eqnarray}
where
${\vec k'}_{\perp}$ is given in Eq. (\ref{BDF1b}),
and
\begin{eqnarray}
A_{\rm b}(q^2)&=&
\left<\Psi^{\uparrow}(P^+,{\vec P_\perp}={\vec q_\perp})
\right|\frac{T^{++}_{\rm b}(0)}{2(P^+)^2}
\left|\Psi^{\uparrow}(P^+,{\vec P_\perp}={\vec
0_\perp})\right>
\nonumber\\
&=&\int\frac{{\mathrm d}^2 {\vec k}_{\perp} {\mathrm d} x }{16 \pi^3}
\ (1-x)\
\Big[\psi^{\uparrow\ *}_{+\frac{1}{2}\, +1}(x,{\vec k''}_{\perp})
\psi^{\uparrow}_{+\frac{1}{2}\, +1}(x,{\vec k}_{\perp})
+\psi^{\uparrow\ *}_{+\frac{1}{2}\, -1}(x,{\vec k''}_{\perp})
\psi^{\uparrow}_{+\frac{1}{2}\, -1}(x,{\vec k}_{\perp})
\nonumber\\
&&\qquad\qquad\qquad\qquad
+\psi^{\uparrow\ *}_{-\frac{1}{2}\, +1}(x,{\vec k''}_{\perp})
\psi^{\uparrow}_{-\frac{1}{2}\, +1}(x,{\vec k}_{\perp})
\Big]\ ,
\label{BDF1ag}
\end{eqnarray}
where
\begin{equation}
{\vec k''}_{\perp}={\vec k}_{\perp}-x{\vec q}_{\perp}\ .
\label{BDF1bg}
\end{equation}
Note that
\begin{equation}
A_{\rm f}(0)+A_{\rm b}(0)=F_1(0) = 1\ ,
\label{BDF1ah}
\end{equation}
which corresponds to the momentum sum rule.

The fermion and boson contributions to the spin-flip matter form factor
are
\begin{eqnarray}
B_{\rm f}(q^2)&=&
{-2M\over (q^1-{\mathrm i}q^2)}
\left<\Psi^{\uparrow}(P^+,{\vec P_\perp}={\vec q_\perp})
\right|\frac{T^{++}_{\rm f}(0)}{2(P^+)^2}
\left|\Psi^{\downarrow}(P^+,{\vec P_\perp}={\vec
0_\perp})\right>
\nonumber\\
&=&{-2M\over (q^1-{\mathrm i}q^2)}
\int\frac{{\mathrm d}^2 {\vec k}_{\perp} {\mathrm d} x }{16 \pi^3}\ x\
\nonumber\\
&&\qquad\qquad\times\Big[\psi^{\uparrow\ *}_{+\frac{1}{2}\, -1}(x,{\vec
k'}_{\perp})
\psi^{\downarrow}_{+\frac{1}{2}\, -1}(x,{\vec k}_{\perp})
+\psi^{\uparrow\ *}_{-\frac{1}{2}\, +1}(x,{\vec k'}_{\perp})
\psi^{\downarrow}_{-\frac{1}{2}\, +1}(x,{\vec k}_{\perp})
\Big]
\nonumber\\
&=&4M\int\frac{{\mathrm d}^2 {\vec k}_{\perp} {\mathrm d} x }{16 \pi^3}
(m-Mx)\varphi (x,{\vec k'}_{\perp}){}^*
\varphi (x,{\vec k}_{\perp})
\nonumber\\
&=&4Me^2\int\frac{{\mathrm d}^2 {\vec k}_{\perp} {\mathrm d} x }{16 \pi^3}
{(m-xM)\over (1-x)}
\nonumber\\
&&\times {1\over [{M^2-(({\vec k}_{\perp}+(1-x){\vec q}_{\perp})^2+m^2)/x
-(({\vec k}_{\perp}+(1-x){\vec q}_{\perp})^2+\lambda^2)/(1-x)}]}
\nonumber\\
&&\times {1\over
[{M^2-({\vec k}_{\perp}^2+m^2)/x-({\vec k}_{\perp}^2+\lambda^2)/(1-x)}]}
\nonumber\\
&=&{Me^2\over 4\pi^2}\int_0^1d\alpha\,\int_0^1dx\,\,
{x(m-xM)\over
\alpha (1-\alpha)\, {1-x\over x}\, {\vec q}_{\perp}^2-M^2
+{m^2\over x}+{\lambda^2\over 1-x}}
\ ,
\label{BDF2abe}
\end{eqnarray}
and
\begin{eqnarray}
B_{\rm b}(q^2)&=&
{-2M\over (q^1-{\mathrm i}q^2)}
\left<\Psi^{\uparrow}(P^+,{\vec P_\perp}={\vec q_\perp})
\right|\frac{T^{++}_{\rm b}(0)}{2(P^+)^2}
\left|\Psi^{\downarrow}(P^+,{\vec P_\perp}={\vec
0_\perp})\right>
\nonumber\\
&=&{-2M\over (q^1-{\mathrm i}q^2)}
\int\frac{{\mathrm d}^2 {\vec k}_{\perp} {\mathrm d} x }{16 \pi^3}\ (1-x)\
\nonumber\\
&&\qquad\qquad\times\Big[\psi^{\uparrow\ *}_{+\frac{1}{2}\, -1}(x,{\vec
k'}_{\perp})
\psi^{\downarrow}_{+\frac{1}{2}\, -1}(x,{\vec k}_{\perp})
+\psi^{\uparrow\ *}_{-\frac{1}{2}\, +1}(x,{\vec k'}_{\perp})
\psi^{\downarrow}_{-\frac{1}{2}\, +1}(x,{\vec k}_{\perp})
\Big]
\nonumber\\
&=&-4M\int\frac{{\mathrm d}^2 {\vec k}_{\perp} {\mathrm d} x }{16 \pi^3}
(m-Mx)\varphi (x,{\vec k'}_{\perp}){}^*
\varphi (x,{\vec k}_{\perp})
\nonumber\\
&=&-4Me^2\int\frac{{\mathrm d}^2 {\vec k}_{\perp} {\mathrm d} x }{16 \pi^3}
{(m-xM)\over (1-x)}
\nonumber\\
&&\times {1\over [{M^2-(({\vec k}_{\perp}-x{\vec q}_{\perp})^2+m^2)/x
-(({\vec k}_{\perp}-x{\vec q}_{\perp})^2+\lambda^2)/(1-x)}]}
\nonumber\\
&&\times {1\over
[{M^2-({\vec k}_{\perp}^2+m^2)/x-({\vec k}_{\perp}^2+\lambda^2)/(1-x)}]}
\nonumber\\
&=&-{Me^2\over 4\pi^2}\int_0^1d\alpha\,\int_0^1dx\,\,
{x(m-xM)\over
\alpha (1-\alpha)\, {x\over 1-x}\, {\vec q}_{\perp}^2-M^2
+{m^2\over x}+{\lambda^2\over 1-x}}
\ .
\label{BDF2abg}
\end{eqnarray}
The total contribution for general momentum transfer is
\begin{eqnarray}
&&B(q^2) = B_{\rm f}(q^2)+B_{\rm b}(q^2)
\nonumber\\
&=&
4Me^2\int\frac{{\mathrm d}^2 {\vec k}_{\perp} {\mathrm d} x }{16 \pi^3}
{(m-xM)\over (1-x)}
\nonumber\\
&&\times
{\{}
{1\over [{M^2-(({\vec k}_{\perp}+(1-x){\vec q}_{\perp})^2+m^2)/x
-(({\vec k}_{\perp}+(1-x){\vec q}_{\perp})^2+\lambda^2)/(1-x)}]}
\nonumber\\
&&\qquad -
{1\over [{M^2-(({\vec k}_{\perp}-x{\vec q}_{\perp})^2+m^2)/x
-(({\vec k}_{\perp}-x{\vec q}_{\perp})^2+\lambda^2)/(1-x)}]} {\}}
\nonumber\\
&&\times
{1\over
[{M^2-({\vec k}_{\perp}^2+m^2)/x-({\vec k}_{\perp}^2+\lambda^2)/(1-x)}]}
\nonumber\\
&=&{Me^2\over 4\pi^2}\int_0^1d\alpha\,\int_0^1dx\,\,
{x(m-xM)}
\label{BDF2abgs}\\
&&\times\Bigl( {1\over
\alpha (1-\alpha)\, {1-x\over x}\, {\vec q}_{\perp}^2-M^2
+{m^2\over x}+{\lambda^2\over 1-x}}
-{1\over
\alpha (1-\alpha)\, {x\over 1-x}\, {\vec q}_{\perp}^2-M^2
+{m^2\over x}+{\lambda^2\over 1-x}}\Bigr)
\ .
\nonumber
\end{eqnarray}
This is the analog of the Pauli form factor for a physical electron
scattering in a gravitational field and in general is not zero.  
However at
zero momentum transfer
\begin{equation}
B(0) = B_{\rm f}(0)+B_{\rm b}(0)=0,
\label{BDF2abgt}
\end{equation}
in agreement with
classical arguments based on the equivalence principle and
conservation of the energy momentum
tensor. \cite{Ji:1998pf,Okun,Kob70,Teryaev}

The helicity-flip electromagnetic and gravitational form factors for the
fluctuations of the electron at one-loop are illustrated in Fig. 1. The
cancellation of the sum of graviton couplings $B(q^2)$ to the constituents
at $q^2 = 0$ is evident.

(a) Helicity-flip Pauli form factor $F_2(q^2)$ in QED.  Notice that
$F_2(0)=1/2$.

(b)  Helicity-flip form factor $B_b(q^2)$ of the graviton coupling
to  the boson (photon) constituent of the electron at one-loop order in
QED.  Notice that $B_b(0) = -1/3$.

(c)   Helicity-flip fermion form factor $B_f(q^2)$ of the graviton
coupling to the fermion constituent at one-loop order in QED.  Notice
that $B_f(0)=1/3$, and thus $B_f(0) + B_b(0)=0.$

(d) Helicity-flip Pauli form factor $F_2(q^2)$ in the Yukawa theory.
Notice that in this case $F_2(0)=3/4$.

(e)  Helicity-flip form factor $B_b(q^2)$ of the graviton coupling
to the boson at one-loop order in the Yukawa theory.  Notice that
$B_b(0)= - 5/12$.

(f)  Helicity-flip fermion form factor $B_f(q^2)$ of the graviton
coupling to the fermion constituent at one-loop order in the Yukawa
theory.  Notice that $B_f(0)=5/12$, and thus $B_f(0) + B_b(0)=0.$
}

\vspace{0.5cm}
\begin{figure}[h!tbp]
\begin{center}
\leavevmode
{\epsfbox{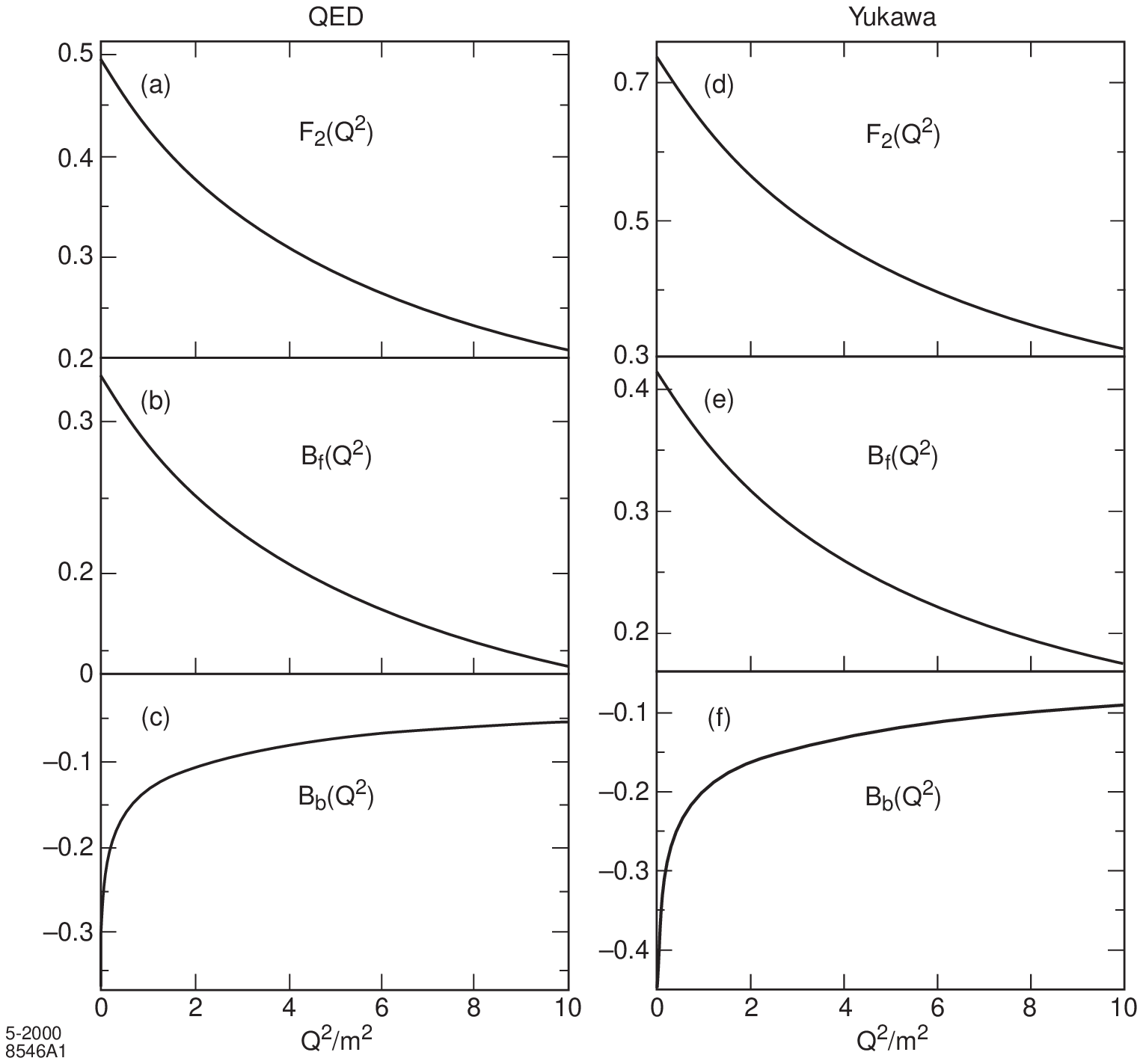}}
\label{FIG1}
\end{center}
\end{figure}

Figure 1: {Helicity-flip electromagnetic and gravitational form
factors for spacelike $q^2=-Q^2<0$
from the quantum fluctuations of a fermion at one-loop order in units of
$\alpha/\pi$ for QED and $g^2/4\pi^2$ for the Yukawa theory.  The fermion
constituent mass is taken as $m_f=M$.  The boson constituent is massless.

In section 5 we shall prove that the anomalous
gravitomagnetic moment $B(0)=B_{\rm f}(0)+B_{\rm b}(0)$ is
identically zero for arbitrary composite systems in quantum field theory.
The proof follows
from the Lorentz covariance of the light-cone
wavefunction and applies to the contribution of each individual Fock
component.

\section{ The Light-Cone Fock State Decomposition and Spin Structure of
Composite Fermions in Yukawa Theory}

As a second example, we shall consider
an effectively composite system
composed of a fermion and a neutral scalar based on the one-loop quantum
fluctuations of this theory.  The light-cone wavefunctions
describe off-shell particles but are
computable explicitly from perturbation theory.
We consider the Yukawa Lagrangian
\begin{eqnarray}
{\cal{L}}&=&{i\over 2}[{\bar{\psi}}\gamma^\mu (\partial_\mu \psi )
-(\partial_\mu {\bar{\psi}})\gamma^\mu\psi]-m{\bar{\psi}}\psi
\label{fmm1}\\[1ex] &+&{1\over 2}(\partial^\mu \phi) (\partial_\mu
\phi) -{1\over 2}\lambda^2\phi\phi
 + g\phi {\bar{\psi}}\psi\ ,
\nonumber
\end{eqnarray}
and the corresponding energy-momentum tensor is given by
\begin{equation}
T^{\mu\nu}= {i\over 2}[{\bar{\psi}}\gamma^\mu
\overrightarrow{\partial}{}^\nu\psi -{\bar{\psi}}\gamma^\mu
\overleftarrow{\partial}{}^\nu\psi ]+ (\partial^\mu\phi)(\partial^\nu\phi)
-g^{\mu\nu}{\cal{L}}\ ,
\label{lags}
\end{equation}
which is conserved.

The $J^z = + {1\over 2}$ two-particle Fock state is given by
\begin{eqnarray}
&&\left|\Psi^{\uparrow}_{\rm two \ particle}(P^+, \vec P_\perp = \vec
0_\perp)\right>
\label{sn1}\\
&=&
\int\frac{{\mathrm d}^2 {\vec k}_{\perp} {\mathrm d} x }{{\sqrt{x(1-x)}}16
\pi^3}
\Big[ \
\psi^{\uparrow}_{+\frac{1}{2}} (x,{\vec k}_{\perp})\,
\left| +\frac{1}{2}\, ;\,\, xP^+\, ,\,\, {\vec k}_{\perp} \right>
+\psi^{\uparrow}_{-\frac{1}{2}} (x,{\vec k}_{\perp})\,
\left| -\frac{1}{2}\, ;\,\, xP^+\, ,\,\, {\vec k}_{\perp} \right>\ \Big]\ ,
\nonumber
\end{eqnarray}
where
\begin{equation}
\left
\{ \begin{array}{l}
\psi^{\uparrow}_{+\frac{1}{2}} (x,{\vec k}_{\perp})=(M+\frac{m}{x})\,
\varphi \ ,\\
\psi^{\uparrow}_{-\frac{1}{2}} (x,{\vec k}_{\perp})=
-\frac{(+k^1+{\mathrm i} k^2)}{x }\,
\varphi \ .
\end{array}
\right.
\label{sn2}
\end{equation}
The scalar part of the wavefunction $\varphi$
is given in Eq. (\ref{wfdenom}) with $e$ replaced by $g$.
The normalization of the Fock states is as in (\ref{normalize}).

Similarly, the $J^z = - {1\over 2}$ two-particle Fock state is given by
\begin{eqnarray}
&&\left|\Psi^{\downarrow}_{\rm two \ particle}(P^+, \vec P_\perp =
\vec 0_\perp)\right>
\label{sn1a}\\
&=&
\int\frac{{\mathrm d}^2 {\vec k}_{\perp} {\mathrm d} x }{{\sqrt{x(1-x)}}16
\pi^3}
\Big[ \
\psi^{\downarrow}_{+\frac{1}{2}} (x,{\vec k}_{\perp})\,
\left| +\frac{1}{2}\, ;\,\, xP^+\, ,\,\, {\vec k}_{\perp} \right>
+\psi^{\downarrow}_{-\frac{1}{2}} (x,{\vec k}_{\perp})\,
\left| -\frac{1}{2}\, ;\,\, xP^+\, ,\,\, {\vec k}_{\perp} \right>\ \Big]\ ,
\nonumber
\end{eqnarray}
where
\begin{equation}
\left
\{ \begin{array}{l}
\psi^{\downarrow}_{+\frac{1}{2}} (x,{\vec k}_{\perp})=
\frac{(+k^1-{\mathrm i} k^2)}{x }\,
\varphi \ ,\\
\psi^{\downarrow}_{-\frac{1}{2}} (x,{\vec k}_{\perp})=(M+\frac{m}{x})\,
\varphi \ .
\end{array}
\right.
\label{sn2a}
\end{equation}
In (\ref{sn2}) and (\ref{sn2a}) we have generalized the framework of
the Yukawa theory by assigning a mass $M$ to the external electrons,
but a different mass $m$ to the internal electron
lines and a mass $\lambda$ to the internal boson line
\cite{BD80}.  The idea behind this is to model the structure
of a composite fermion state with mass $M$ by a fermion and a boson
constituent with respective masses $m$ and $\lambda$.

Using Eqs. (\ref{BD1}) and (\ref{sn1}) we
have
\begin{equation}
F_1(q^2) = Z +
\int\frac{{\mathrm d}^2 {\vec k}_{\perp} {\mathrm d} x }{16 \pi^3}
\Big[\psi^{\uparrow\ *}_{+\frac{1}{2}}(x,{\vec k'}_{\perp})
\psi^{\uparrow}_{+\frac{1}{2}}(x,{\vec k}_{\perp})
+\psi^{\uparrow\ *}_{-\frac{1}{2}}(x,{\vec k'}_{\perp})
\psi^{\uparrow}_{-\frac{1}{2}}(x,{\vec k}_{\perp})\Big]\ ,
\label{BDF1as}
\end{equation}
where ${\vec k'}_{\perp}$ is given in Eq. (\ref{BDF1b}).
At zero momentum transfer
\begin{equation}
F_1(0) = Z +
\int\frac{{\mathrm d}^2 {\vec k}_{\perp} {\mathrm d} x }{16 \pi^3}
\Big[ \psi^{\uparrow\ *}_{+\frac{1}{2}}(x,{\vec k}_{\perp})
\psi^{\uparrow}_{+\frac{1}{2}}(x,{\vec k}_{\perp})
+\psi^{\uparrow\ *}_{-\frac{1}{2}}(x,{\vec k}_{\perp})
\psi^{\uparrow}_{-\frac{1}{2}}(x,{\vec k}_{\perp})\Big]\ ,
\label{BDF1s}
\end{equation}
where the renormalization constant $Z$ in light-cone gauge is given, using
Pauli-Villars regularization, by
\begin{eqnarray}
Z = 1 &-& {1\over 4\pi}({g^2\over 4\pi})\int_0^1 dx\ \Big[ \
(1-x)\,\, {\rm ln}{-M^2+{m^2\over x}+{\Lambda^2\over 1-x}\over
-M^2+{m^2\over x}+{\lambda^2\over 1-x}} \nonumber\\ && +\
{(xM+m)^2\over x}\, \Bigl( -{1\over -M^2+{m^2\over
x}+{\Lambda^2\over 1-x}} +{1\over -M^2+{m^2\over
x}+{\lambda^2\over 1-x}}\Bigr) \ \Big] \ .
\end{eqnarray}

The Pauli form factor is obtained from the spin-flip matrix element of
the $J^+$ current.  From Eqs. (\ref{BD1}), (\ref{sn1}) and (\ref{sn1a}) we
have
\begin{eqnarray}
F_2(q^2)
&=&{-2M\over (q^1-{\mathrm i}q^2)}
\int\frac{{\mathrm d}^2 {\vec k}_{\perp} {\mathrm d} x }{16 \pi^3}
\Big[\psi^{\uparrow\ *}_{+\frac{1}{2}}(x,{\vec k'}_{\perp})
\psi^{\downarrow}_{+\frac{1}{2}}(x,{\vec k}_{\perp})
+\psi^{\uparrow\ *}_{-\frac{1}{2}}(x,{\vec k'}_{\perp})
\psi^{\downarrow}_{-\frac{1}{2}}(x,{\vec k}_{\perp})
\Big]
\nonumber\\
&=&2M\int\frac{{\mathrm d}^2 {\vec k}_{\perp} {\mathrm d} x }{16 \pi^3}
{(1-x)(m+Mx)\over x^2}\varphi (x,{\vec k'}_{\perp}){}^*
\varphi (x,{\vec k}_{\perp})
\nonumber\\
&=&2Mg^2\int\frac{{\mathrm d}^2 {\vec k}_{\perp} {\mathrm d} x }{16 \pi^3}
{(m+xM)\over x^2}
\nonumber\\
&&\times {1\over [{M^2-(({\vec k}_{\perp}+(1-x){\vec q}_{\perp})^2+m^2)/x
-(({\vec k}_{\perp}+(1-x){\vec q}_{\perp})^2+\lambda^2)/(1-x)}]}
\nonumber\\
&&\times {1\over
[{M^2-({\vec k}_{\perp}^2+m^2)/x-({\vec k}_{\perp}^2+\lambda^2)/(1-x)}]}
\nonumber\\
&=&{Mg^2\over 8\pi^2}\int_0^1d\alpha\,\int_0^1dx\,\,
{{1-x\over x}(m+xM)\over
\alpha (1-\alpha)\, {1-x\over x}\, {\vec q}_{\perp}^2-M^2
+{m^2\over x}+{\lambda^2\over 1-x}}
\ .
\label{BDF2as}
\end{eqnarray}
The anomalous moment is obtained in the limit of zero momentum transfer,
\begin{eqnarray}
F_2(0)&=&
2Mg^2\int\frac{{\mathrm d}^2 {\vec k}_{\perp} {\mathrm d} x }{16 \pi^3}
{(m+xM) \over x^2}\,\, {1\over
[{M^2-({\vec k}_{\perp}^2+m^2)/x-({\vec k}_{\perp}^2+\lambda^2)/(1-x)}]^2}
\nonumber\\
&=&{Mg^2\over 8\pi^2}\int_0^1dx\,\,
{{1-x\over x}(m+xM)\over
-M^2
+{m^2\over x}+{\lambda^2\over 1-x}}
\ .
\label{BDF2s}
\end{eqnarray}

The individual contributions of the fermion and boson fields
to the energy-momentum form factors in the Yukawa model are given by
\begin{eqnarray}
A_{\rm f}(q^2)&=&
\left<\Psi^{\uparrow}(P^+,{\vec P_\perp}={\vec q_\perp})
\right|x\left|\Psi^{\uparrow}(P^+,{\vec P_\perp}={\vec 0_\perp})\right>
\label{BDF1aes}\\
&=&\int\frac{{\mathrm d}^2 {\vec k}_{\perp} {\mathrm d} x }{16 \pi^3}
\ x\
\Big[\psi^{\uparrow\ *}_{+\frac{1}{2}}(x,{\vec k'}_{\perp})
\psi^{\uparrow}_{+\frac{1}{2}}(x,{\vec k}_{\perp})
+\psi^{\uparrow\ *}_{-\frac{1}{2}}(x,{\vec k'}_{\perp})
\psi^{\uparrow}_{-\frac{1}{2}}(x,{\vec k}_{\perp})\Big]\ ,
\nonumber
\end{eqnarray}
where
$ {\vec k'}_{\perp}={\vec k}_{\perp}+(1-x){\vec q}_{\perp} ,$
and
\begin{eqnarray}
A_{\rm b}(q^2)&=&
\left<\Psi^{\uparrow}(P^+,{\vec P_\perp}={\vec q_\perp})
\right|(1-x)\left|\Psi^{\uparrow}(P^+,{\vec P_\perp}={\vec
0_\perp})\right>
\label{BDF1ags}\\
&=&\int\frac{{\mathrm d}^2 {\vec k}_{\perp} {\mathrm d} x }{16 \pi^3}
\ (1-x)\
\Big[\psi^{\uparrow\ *}_{+\frac{1}{2}}(x,{\vec k''}_{\perp})
\psi^{\uparrow}_{+\frac{1}{2}}(x,{\vec k}_{\perp})
+\psi^{\uparrow\ *}_{-\frac{1}{2}}(x,{\vec k''}_{\perp})
\psi^{\uparrow}_{-\frac{1}{2}}(x,{\vec k}_{\perp})\Big]\ ,
\nonumber
\end{eqnarray}
where in this case
${\vec k''}_{\perp}={\vec k}_{\perp}-x{\vec q}_{\perp} .$
Note that again
\begin{equation}
A_{\rm f}(0)+A_{\rm b}(0)=F_1(0) = 1\ ,
\label{BDF1ahs}
\end{equation}
which corresponds to the momentum sum rule.

The fermion and boson contributions to the spin-flip matter form factor
are
\begin{eqnarray}
B_{\rm f}(q^2)
&=&{-2M\over (q^1-{\mathrm i}q^2)}
\int\frac{{\mathrm d}^2 {\vec k}_{\perp} {\mathrm d} x }{16 \pi^3}\ x\
\nonumber\\
&&\qquad\qquad\times\
\Big[\psi^{\uparrow\ *}_{+\frac{1}{2}}(x,{\vec k'}_{\perp})
\psi^{\downarrow}_{+\frac{1}{2}}(x,{\vec k}_{\perp})
+\psi^{\uparrow\ *}_{-\frac{1}{2}}(x,{\vec k'}_{\perp})
\psi^{\downarrow}_{-\frac{1}{2}}(x,{\vec k}_{\perp})
\Big]
\nonumber\\
&=&2M\int\frac{{\mathrm d}^2 {\vec k}_{\perp} {\mathrm d} x }{16 \pi^3}
{(1-x)(m+Mx)\over x}\varphi (x,{\vec k'}_{\perp}){}^*
\varphi (x,{\vec k}_{\perp})
\nonumber\\
&=&2Mg^2\int\frac{{\mathrm d}^2 {\vec k}_{\perp} {\mathrm d} x }{16 \pi^3}
{(m+xM)\over x}
\nonumber\\
&&\times {1\over [{M^2-(({\vec k}_{\perp}+(1-x){\vec q}_{\perp})^2+m^2)/x
-(({\vec k}_{\perp}+(1-x){\vec q}_{\perp})^2+\lambda^2)/(1-x)}]}
\nonumber\\
&&\times {1\over
[{M^2-({\vec k}_{\perp}^2+m^2)/x-({\vec k}_{\perp}^2+\lambda^2)/(1-x)}]}
\nonumber\\
&=&{Mg^2\over 8\pi^2}\int_0^1d\alpha\,\int_0^1dx\,\,
{(1-x)(m+xM)\over
\alpha (1-\alpha)\, {1-x\over x}\, {\vec q}_{\perp}^2-M^2
+{m^2\over x}+{\lambda^2\over 1-x}}
\ .
\label{BDF2abes}
\end{eqnarray}
and
\begin{eqnarray}
B_{\rm b}(q^2)
&=&{-2M\over (q^1-{\mathrm i}q^2)}
\int\frac{{\mathrm d}^2 {\vec k}_{\perp} {\mathrm d} x }{16 \pi^3}\ (1-x)\
\nonumber\\
&&\qquad\qquad\times\
\Big[\psi^{\uparrow\ *}_{+\frac{1}{2}\, -1}(x,{\vec k'}_{\perp})
\psi^{\downarrow}_{+\frac{1}{2}\, -1}(x,{\vec k}_{\perp})
+\psi^{\uparrow\ *}_{-\frac{1}{2}\, +1}(x,{\vec k'}_{\perp})
\psi^{\downarrow}_{-\frac{1}{2}\, +1}(x,{\vec k}_{\perp})
\Big]
\nonumber\\
&=&-2M\int\frac{{\mathrm d}^2 {\vec k}_{\perp} {\mathrm d} x }{16 \pi^3}
{(1-x)(m+Mx)\over x}\varphi (x,{\vec k''}_{\perp}){}^*
\varphi (x,{\vec k}_{\perp})
\nonumber\\
&=&-2Mg^2\int\frac{{\mathrm d}^2 {\vec k}_{\perp} {\mathrm d} x }{16 \pi^3}
{(m+xM)\over x}
\nonumber\\
&&\times {1\over [{M^2-(({\vec k}_{\perp}-x{\vec q}_{\perp})^2+m^2)/x
-(({\vec k}_{\perp}-x{\vec q}_{\perp})^2+\lambda^2)/(1-x)}]}
\nonumber\\
&&\times {1\over
[{M^2-({\vec k}_{\perp}^2+m^2)/x-({\vec k}_{\perp}^2+\lambda^2)/(1-x)}]}
\nonumber\\
&=&-{Mg^2\over 8\pi^2}\int_0^1d\alpha\,\int_0^1dx\,\,
{(1-x)(m+xM)\over
\alpha (1-\alpha)\, {x\over 1-x}\, {\vec q}_{\perp}^2-M^2
+{m^2\over x}+{\lambda^2\over 1-x}}
\ .
\label{BDF2abgss}
\end{eqnarray}

The total contribution from the fermion and boson constituents is
\begin{eqnarray}
&&B(q^2) = B_{\rm f}(q^2)+B_{\rm b}(q^2)
\nonumber\\
&=&
2Mg^2\int\frac{{\mathrm d}^2 {\vec k}_{\perp} {\mathrm d} x }{16 \pi^3}
{(m+xM)\over x}
\nonumber\\
&&\times
{\{}
{1\over [{M^2-(({\vec k}_{\perp}+(1-x){\vec q}_{\perp})^2+m^2)/x
-(({\vec k}_{\perp}+(1-x){\vec q}_{\perp})^2+\lambda^2)/(1-x)}]}
\nonumber\\
&&\qquad -
{1\over [{M^2-(({\vec k}_{\perp}-x{\vec q}_{\perp})^2+m^2)/x
-(({\vec k}_{\perp}-x{\vec q}_{\perp})^2+\lambda^2)/(1-x)}]} {\}}
\nonumber\\
&&\times
{1\over
[{M^2-({\vec k}_{\perp}^2+m^2)/x-({\vec k}_{\perp}^2+\lambda^2)/(1-x)}]}
\nonumber\\
&=&{Mg^2\over 8\pi^2}\int_0^1d\alpha\,\int_0^1dx\,\,
{(1-x)(m+xM)}
\label{BDF2abgssa}\\
&&\times \Bigl( {1\over
\alpha (1-\alpha)\, {1-x\over x}\, {\vec q}_{\perp}^2-M^2
+{m^2\over x}+{\lambda^2\over 1-x}}
-{1\over
\alpha (1-\alpha)\, {x\over 1-x}\, {\vec q}_{\perp}^2-M^2
+{m^2\over x}+{\lambda^2\over 1-x}}\Bigr)
\ .
\nonumber
\end{eqnarray}
At
zero momentum transfer
\begin{equation}
B(0) = B_{\rm f}(0)+B_{\rm b}(0)=0 \ ,
\label{BDF2abgts}
\end{equation}
which is another example of the vanishing of the anomalous gravitomagnetic
moment.  [See also Fig. 1(e) + Fig. 1(f).]
The general proof that $B(0) = 0 $ for any system is given in the next
section.  Note that $B(Q^2)$ does not vanish for nonzero momentum
transfer.

\section {The Anomalous Gravitomagnetic Moment for Composite Systems}

In this section we shall show that the anomalous gravitomagnetic
moment $B(0)$ always vanishes for each contributing Fock state of a
general composite system.  In order to calculate $B(0)$ using Eq.
(\ref{eBD2}), we need to consider a non-forward amplitude.
The internal momentum variables for the final state wavefunction are
given by Eqs. (\ref{kprime1}) and (\ref{kprime2}).
The subscripts of $x_i$ and $\vec k_{\perp i}$
label constituent particles, the superscripts of $q^1_\perp$, $k^1_\perp$,
and $k^2_\perp$ label the Lorentz indices, and the subscript $a$ in
$\psi_a$ indicates the contributing Fock state.  The essential ingredient
is the Lorentz property of the light-cone wavefunctions.

It is important to identify the $n-1$ independent relative momenta of
the $n$-particle Fock state.
\begin{eqnarray}
&&-{B(0)\over 2M}
=
\lim_{q_{\perp}^1\to 0}{\partial\over \partial q_{\perp}^1}
\VEV{P+q,\uparrow\left|\frac{T^{++}(0)}{2(P^+)^2}\right|P,\downarrow}
\label{gravito1}\\
&=&\lim_{q_{\perp}^1\to 0}{\partial\over \partial q_{\perp}^1}
\left<\Psi^{\uparrow}(P^+=1, {\vec P_\perp}= \vec
q_\perp))\left|\frac{T^{++}(0)}{2(P^+)^2}
\right|\Psi^{\downarrow}(P^+=1, {\vec
P_\perp}= \vec 0_\perp)\right>
\nonumber\\
&=&\lim_{q_{\perp}^1\to 0}{\partial\over \partial q_{\perp}^1}
\sum_{a}\int\prod^{n-1}_{k=1}
\frac{{\mathrm d}^2 {\vec k}_{\perp k} {\mathrm d} x_k }{16 \pi^3}
\psi^{\uparrow *}_a\Big( x_1,x_2,\cdots ,x_{n-1},(1-x_1-x_2-\cdots -x_{n-1}),
\nonumber\\
&&\qquad\qquad
{\vec k}_{\perp 1}',{\vec k}_{\perp 2}',\cdots ,{\vec k}_{\perp n-1}',
(-{\vec k}_{\perp 1}'-{\vec k}_{\perp 2}'-\cdots -{\vec k}_{\perp n-1}')\Big)
\nonumber\\
&\times &\Big[ \sum_{i=1}^{n-1}x_i+(1-x_1-x_2-\cdots -x_{n-1})\Big]
\nonumber\\
&\times &\psi^{\downarrow}_a
\Big( x_1,x_2,\cdots ,x_{n-1},(1-x_1-x_2-\cdots -x_{n-1}),
\nonumber\\
&&\qquad\qquad
{\vec k}_{\perp 1},{\vec k}_{\perp 2},\cdots ,{\vec k}_{\perp n-1},
(-{\vec k}_{\perp 1}-{\vec k}_{\perp 2}-\cdots
-{\vec k}_{\perp n-1})\Big) \ .
\nonumber
\end{eqnarray}

Using integration by parts,
\begin{eqnarray}
&&-{B_a(0)\over 2M}
=
\label{gravito2}\\
&=&\int\prod^{n-1}_{k=1}
\frac{{\mathrm d}^2 {\vec k}_{\perp k} {\mathrm d} x_k }{16 \pi^3}
\psi^{\uparrow *}_a\Big( x_1,x_2,\cdots ,x_{n-1},(1-x_1-x_2-\cdots -x_{n-1}),
\nonumber\\
&&\qquad\qquad
{\vec k}_{\perp 1},{\vec k}_{\perp 2},\cdots ,{\vec k}_{\perp n-1},
(-{\vec k}_{\perp 1}-{\vec k}_{\perp 2}-\cdots -{\vec k}_{\perp n-1})\Big)
\nonumber\\
&\times &
\Big[
\sum_{i=1}^{n-1}x_i\Big( (-1+x_i){\partial\over\partial k_{\perp i}^1}+
\sum_{j \ne i}^{n-1}x_j{\partial\over\partial k_{\perp j}^1}\Big)
+(1-x_1-x_2-\cdots -x_{n-1})
\sum_{j=1}^{n-1}x_j{\partial\over\partial k_{\perp j}^1}\Big]
\nonumber\\
&\times &\psi^{\downarrow}_a
\Big( x_1,x_2,\cdots ,x_{n-1},(1-x_1-x_2-\cdots -x_{n-1}),
\nonumber\\
&&\qquad\qquad
{\vec k}_{\perp 1},{\vec k}_{\perp 2},\cdots ,{\vec k}_{\perp n-1},
(-{\vec k}_{\perp 1}-{\vec k}_{\perp 2}-\cdots -{\vec k}_{\perp n-1})\Big)
\nonumber\\
&=&\int\prod^{n-1}_{k=1}
\frac{{\mathrm d}^2 {\vec k}_{\perp k} {\mathrm d} x_k }{16 \pi^3}
\psi^{\uparrow *}_a\Big( x_1,x_2,\cdots ,x_{n-1},(1-x_1-x_2-\cdots -x_{n-1}),
\nonumber\\
&&\qquad\qquad
{\vec k}_{\perp 1},{\vec k}_{\perp 2},\cdots ,{\vec k}_{\perp n-1},
(-{\vec k}_{\perp 1}-{\vec k}_{\perp 2}-\cdots -{\vec k}_{\perp n-1})\Big)
\nonumber\\
&\times &
\Big[
\sum_{i=1}^{n-1}
\Big( -1+\sum_{j=1}^{n-1}x_j+(1-x_1-x_2-\cdots -x_{n-1})\Big)
x_i{\partial\over\partial k_{\perp i}^1}\Big]
\nonumber\\
&\times &\psi^{\downarrow}_a
\Big( x_1,x_2,\cdots ,x_{n-1},(1-x_1-x_2-\cdots -x_{n-1}),
\nonumber\\
&&\qquad\qquad
{\vec k}_{\perp 1},{\vec k}_{\perp 2},\cdots ,{\vec k}_{\perp n-1},
(-{\vec k}_{\perp 1}-{\vec k}_{\perp 2}-\cdots -{\vec k}_{\perp n-1})\Big)
\nonumber\\
&=&0\ .
\nonumber
\end{eqnarray}
Thus the contribution $B_a(0)$ from each contributing Fock state $a$ to
the total anomalous gravitomagnetic moment $B(0)$ vanishes separately.

\section {The Perturbative Models as a Template for a Composite System}

We can use the structure of the one-loop QED and Yukawa calculations with
general values for $M$,
$m$, and
$\lambda$, to represent a spin-$1\over 2$ system composed of a fermion
and a spin-1 or spin-0 boson.  Such a model describes an effectively
composite system with no bare one-particle Fock state.  We can also
generalize the functional form of the momentum space wavefunction
$\varphi(x,\vec k_\perp)$ by introducing a spectrum of vector
bosons satisfying the generalized Pauli-Villars spectral conditions
\begin{equation}
\int d \lambda^2  \lambda^{2N} \rho(\lambda^2) = 0, \ \ \ N= 0, 1, \cdots
\ .
\label{BDF2y}
\end{equation}
For example, we can simulate a proton
as a bound state of a quark and diquark \cite{Close}, using
spin-0, spin-1 diquarks, or a linear
superposition of the two states.
The model can be made to match the power-law fall-off of the hadron form
factors predicted in perturbative QCD by the choice of
sum rule conditions on the Pauli-Villars spectra.\cite{Bro92,
Brodsky:1998hs}.  The light-cone framework of the model
resembles that of the covariant parton model of Landshoff, Polkinghorne
and Short
\cite{Landshoff:1971ff, Brodsky:1973hm}, in which the power behavior of
the spectral integral at high masses corresponds to the Regge behavior of
the deep inelastic structure functions.
Although the model is based on just two Fock constituents,
it is relativistic and satisfies self-consistency
conditions such as in the point-like limit where
$R^2 M^2 \to 0$, the anomalous moment vanishes.\cite{Bro94}
The light-cone formalism
also properly incorporates Wigner boosts.  Thus this model of composite
systems can serve as a useful theoretical laboratory to interrelate
hadronic properties and check the consistency of formulae proposed for the
study of hadron substructure.

\section{ Spin and Orbital Angular Momentum Composition of Light-Cone
Wavefunctions}

In general the light-cone wavefunctions satisfy conservation of the
$z$ projection of angular momentum:
\begin{equation}
J^z = \sum^n_{i=1} s^z_i + \sum^{n-1}_{j=1} l^z_j \ .
\label{Jzsumrule}
\end{equation}
The sum over $s^z_i$
represents the contribution of the intrinsic spins of the $n$ Fock state
constituents.  The sum over orbital angular momenta
$l^z_j = -{\mathrm i} (k^1_j\frac{\partial}{\partial k^2_j}
-k^2_j\frac{\partial}{\partial k^1_j})$ derives from
the $n-1$ relative momenta.  This excludes the contribution to the
orbital angular momentum due to the motion of the center of mass, which
is not an intrinsic property of the hadron.

We can see
how the angular momentum sum rule Eq. (\ref{Jzsumrule}) is satisfied for
the wavefunctions Eqs. (\ref{vsn1}) and (\ref{vsn1a}) of the QED model
system  of two-particle Fock states.  In Table~1 we list the fermion
constituent's light-cone spin projection
$s^z_{\rm f} = {1\over 2} \lambda_{\rm f}$, the boson constituent spin
projection
$s^z_{\rm b} = \lambda_{\rm b}$, and the relative orbital angular momentum
$l^z$ for each contributing configuration of the QED model system
wavefunction.
\begin{table}[ht]
\begin{center}
Table 1.  Spin Decomposition of the $J^z_e = + 1/2$ Electron  \\
\vspace{.6truecm}
\begin{footnotesize}
\begin{tabular}{|c|c|c|c|}
\hline\ru1
$\ \ \ \ $Configuration$\ \ \ \ $ & $\ \ \ $Fermion Spin $s^z_{\rm f}$$\ \ \ $
&$\ \ \ $ Boson Spin $s^z_{\rm b}$$\ \ \ $
& Orbital Ang.  Mom. $l^z$\\
\hline\hline\ru1
{$\left|+\frac{1}{2}\right> \to  \left|+\frac{1}{2}\, +1\right> $}
& {$+{1\over 2}$}
& {$+1$}
& {$-1$}\\
\hline\ru1
{$\left|+\frac{1}{2}\right> \to  \left|-\frac{1}{2}\, +1\right> $}
& {$-{1\over 2}$}
& {$+1$}
& {$0$}\\
\hline\ru1
{$\left|+\frac{1}{2}\right> \to  \left|+\frac{1}{2}\, -1\right> $}
& {$+{1\over 2}$}
& {$-1$}
& {$+1$}\\
\hline
\end{tabular}
\end{footnotesize}
\end{center}
\end{table}
Table~1 
is derived by
calculating the matrix elements of the light-cone helicity
operator $\gamma^+\gamma^5$ \cite{Ma91b}
and the relative orbital angular momentum
operator $-{\mathrm i} (k^1\frac{\partial}{\partial k^2}
-k^2\frac{\partial}{\partial k^1})$ \cite{Harindranath:1999ve,MS98,Hag98}
in the light-cone representation. Each configuration satisfies
the spin sum rule: $J^z=s^z_{\rm f}+s^z_{\rm b} + l^z$.

For a better understanding of Table~1, we look at the
non-relativistic and ultra-relativistic limits.
At the non-relativistic limit, the transversal motions
of the constituent can be neglected and we have only
the $\left|+\frac{1}{2}\right> \to \left|-\frac{1}{2}\, +1\right>$
configuration which is the non-relativistic quantum
state for the spin-half system composed of
a fermion and a spin-1 boson constituents. The fermion
constituent has spin projection in the opposite
direction to the spin $J^z$ of the whole system.
However, for  ultra-relativistic binding  in which
the transversal motions of the constituents are large compared to the
fermion masses,  the
$\left|+\frac{1}{2}\right> \to \left|+\frac{1}{2}\, +1\right>$
and
$\left|+\frac{1}{2}\right> \to \left|+\frac{1}{2}\, -1\right>$
configurations dominate
over the $\left|+\frac{1}{2}\right> \to \left|-\frac{1}{2}\, +1\right>$
configuration. In this case
the fermion constituent has
spin projection  parallel to $J^z$.

\begin{table}[ht]
\begin{center}
Table 2. Spin Decomposition of the $J^z = + 1/2$ Fermion in Yukawa Theory \\
\vspace{.6truecm}
\begin{footnotesize}
\begin{tabular}{|c|c|c|c|}
\hline\ru1
$\ \ \ \ $Configuration$\ \ \ \ $ & $\ \ \ $Fermion Spin $s^z_{\rm f}$$\ \ \ $
&$\ \ \ $ Boson Spin $s^z_{\rm b}$$\ \ \ $
& Orbital Ang.  Mom. $l^z$\\
\hline\hline\ru1
{$\left|+\frac{1}{2}\right> \to  \left|+\frac{1}{2}\right> $}
& {$+{1\over 2}$}
& {$0$}
& {$0$}\\
\hline\ru1
{$\left|+\frac{1}{2}\right> \to  \left|-\frac{1}{2}\right> $}
& {$-{1\over 2}$}
& {$0$}
& {$+1$}\\
\hline
\end{tabular}
\end{footnotesize}
\end{center}
\end{table}

The corresponding spin content in the Yukawa theory is given in Table 2.
In this case, the non-relativistic fermion's spin projection is aligned
with the total $J^z$, and it is anti-aligned in the
ultra-relativistic limit.
The distinct features of spin structure
in the non-relativistic and ultra-relativistic
limits reveals the importance
of relativistic effects and
supports the viewpoint \cite{Ma91b,Ma96,MSS97}
that the proton ``spin
puzzle" can be understood as due to the relativistic
motion of quarks inside the nucleon.  In particular, the spin projection
of the relativistic constituent quark tends to be anti-aligned with the
proton spin in a quark-diquark bound state if the diquark has spin 0. The
state with orbital angular momentum $l^z= \pm 1 $ in fact dominates over
the states with $l^z = 0.$   Thus the empirical fact that $\Delta q$ is
small in the proton has a natural description in the light-cone Fock
representation of hadrons.

The explicit formulas for the quark spin distributions $\Delta
q(x,\Lambda^2)$ in the quark-diquark models can be immediately obtained
for the spin-1 diquark model from Eqs. (\ref{vsn1}) and (\ref{vsn2}):
\begin{eqnarray}
&&\Delta q(x,\Lambda^2)_{\rm spin-1\ diquark}
\label{dqvector} \\
&=&
\int\frac{{\mathrm d}^2 {\vec k}_{\perp} {\mathrm d} x }{16 \pi^3}
\theta (\Lambda^2 - {\cal M}^2)
\ 2\ \Big[ \ {{\vec k}_{\perp}^2\over x^2(1-x)^2}\
+\ {{\vec k}_{\perp}^2\over (1-x)^2}\
-\ (M-{m\over x})^2\ \Big]\ |\varphi |^2\ ,
\nonumber
\end{eqnarray}
and for the spin-0 diquark model from Eqs. (\ref{sn1}) and (\ref{sn2}):
\begin{equation}
\Delta q(x,\Lambda^2)_{\rm spin-0\ diquark}=
\int\frac{{\mathrm d}^2 {\vec k}_{\perp} {\mathrm d} x }{16 \pi^3}
\theta (\Lambda^2 - {\cal M}^2)
\ \Big[ \
(M+{m\over x})^2\ -\ {{\vec k}_{\perp}^2\over x^2}\ \Big]\ |\varphi |^2\ ,
\label{dqscalar}
\end{equation}
where we have regulated the integral by assuming a cutoff in the
invariant mass: ${\cal M}^2 =
\sum_i {{\vec k}^2_{\perp i} +m^2_i\over x_i} <
\Lambda^2.$
Again, one sees the transition of $\Delta q$  from the
nonrelativistic to relativistic limit.  In the spin-0 diquark model
$\Delta q = 1$ in the nonrelativistic limit, and decreases toward $\Delta
q =  - 1$  as the intrinsic transverse momentum increases. The behavior is
just opposite in the case of the spin-1 diquark.

\section{Conclusions}

The LC wavefunctions $\psi_{n/H}(x_i,\vec
k_{\perp i},\lambda_i)$ provide a general representation of a
relativistic composite system. They are universal, process independent,
and control all hadronic reactions.  In this paper we have
constructed explicit models which are simple but yet are
completely relativistic, preserve all of the Lorentz properties of a
composite system of quantum field theory.  Because of this explicit
realization we can see how different hadronic phenomena can be
interrelated.  For example, the matrix elements of local operators such
as the electromagnetic current, the energy momentum tensor, angular
momentum, and the moments of structure functions have exact
representations in terms of light-cone Fock state wavefunctions of bound
states such as hadrons.  We have shown that each Fock state of a
composite system satisfies $J_z$ conservation, component by component.
We have emphasized that the correct expression for the orbital angular
momentum $l^z$ involves a sum over $n-1$ relative momentum
contributions for a Fock state with $n$ constituents.

We have illustrated these properties by examining the  explicit form
of the light-cone wavefunctions for the
two-particle Fock state of the electron in QED, thus connecting the
Schwinger anomalous magnetic moment to the spin and orbital momentum
carried by its Fock state constituents. We have also computed the QED
one-loop radiative corrections for the form factors for the graviton
coupling to the electron and photon.  The one-loop model provides  a
transparent basis for understanding the structure of
relativistic composite systems and their matrix elements in hadronic
physics.
Although the underlying model is derived from elementary  QED perturbative
couplings, it in fact  can be used to model much more general bound state
systems by applying spectral integration over the constituent masses
while preserving all of the Lorentz properties.

We thus have obtained a theoretical laboratory to
test the consistency of formulae which have been proposed to probe the
spin structure of hadrons. For example, we have computed
the quark spin distributions
$\Delta q(x,\Lambda^2)$ in  quark-diquark models. In particular,
the spin projection of the relativistic constituent quark tends to be
anti-aligned with the proton spin in a quark-diquark bound state if the
diquark has spin 0. The empirical fact that $\Delta q$ is small in the
proton thus has a  natural description in the light-cone Fock
representation of a relativistic bound state.

We  have also given general exact expressions for the matrix elements of
the electromagnetic, electroweak, and graviton couplings for
operators for arbitrary composite systems, giving explicit realization of
the spin sum rules and other local matrix elements in terms of the
light-cone Fock state wavefunctions.

Finally, we  have given a general proof demonstrating that the
anomalous gravitomagnetic moment  $B(0)$ for gravitons coupling to matter
vanishes identically for any composite system.  At one loop order in
QED, we can see the explicit cancellation of the graviton coupling to
the lepton and photon. In fact we have shown that this remarkable
property holds generally for any composite or elementary system
at all orders directly from the Lorentz boost properties of the light-cone
Fock representation.

\bigskip

\end{document}